\documentclass[12pt]{article}
\usepackage{graphicx}
\usepackage{amsmath,amsfonts,amssymb}
\usepackage[margin=1in]{geometry}
\usepackage{hyperref}
\usepackage{float}
\usepackage[T1]{fontenc}

\title{Native quantum games from interacting discrete-time quantum walks}

\author{
    Rashid Ahmad\\
    \small Department of Physics, College of Integrative Studies\\
    \small Abdullah Al Salem University, Kuwait\\
    \small \texttt{rashid.ahmad@aasu.edu.kw}
}

\date{\today}

\begin{document}

\maketitle
\begin{abstract}
We study how strategic interaction can arise from controlled quantum dynamics rather than being imposed as an external mathematical structure. We introduce a class of interaction-defined quantum games in which players are represented by distinguishable quantum walkers, strategies correspond to local coin operations, and payoffs are defined as expectation values of physical observables. Using interacting discrete-time quantum walks as a concrete platform, we demonstrate numerically that competitive, cooperative, and asymmetric games admit stable stationary strategy profiles when the walkers are coupled, while no non-trivial equilibria exist in the absence of interaction. To clarify the game-theoretic structure, we derive an analytic perturbative decomposition of the payoff function in the weak-interaction regime, showing explicitly that strategic coupling originates from interaction-induced interference terms in the joint probability distribution. For a collision-based phase interaction, the payoff becomes non-separable at first order in the interaction strength and generically admits stationary points satisfying the Nash conditions. Our results provide a physically explicit realization of strategic interdependence in quantum transport processes and establish interacting quantum walks as a minimal platform for studying game-theoretic behavior emerging from unitary dynamics.
\end{abstract}


\section{Introduction}

Game theory provides a general language for describing strategic interactions between competing or cooperating agents. In quantum information science, this framework has primarily been explored through the quantization of classical games, most notably in the Eisert--Wilkens--Lewenstein (EWL) construction and its subsequent extensions, where predefined payoff matrices are embedded into Hilbert-space dynamics via entanglement and measurement procedures \cite{Eisert1999,Benjamin2001}. Related formulations include the Marinatto--Weber scheme and continuous-strategy generalizations, which likewise retain an explicit classical payoff table as an external object mapped onto quantum operations \cite{Marinatto2000,Flitney2002}. A general Hilbert-space formulation of quantum games demonstrated that two-strategy quantum games are equivalent to families of classical games augmented by quantum interference, clarifying how quantum strategies can outperform classical ones and revealing phenomena such as phase-dependent cyclic subgames \cite{Cheon2006}. Two-qubit quantum computations have also been modeled as strictly competitive games, where the geometry of Hilbert space establishes the equivalence between Nash equilibrium and mini-max outcomes \cite{Khan2013}. Further studies have highlighted non-classical features arising from unitary strategy spaces and entanglement, as well as applications to opinion formation and economic dynamics \cite{Frackiewicz2021,2016EL11450012,Piotrowski2003}.

Collectively, these approaches demonstrate the power of quantum formalism in extending classical game theory; however, they share a common feature in that the strategic structure is externally specified rather than arising intrinsically from the underlying physical dynamics. Recent developments have extended quantum game theory toward multi-agent and dynamical settings, where entanglement and quantum correlations influence collective behavior and long-term dynamics \cite{Teeni2023,Shi2025}.

Experimental realizations of such quantum games have been demonstrated on several quantum information platforms, including nuclear magnetic resonance processors, linear optics, and superconducting qubits \cite{Du2002,Prevedel2007}. These implementations confirm the operational viability of quantum game protocols, but they also highlight that the strategic structure is typically imposed at the level of circuit design rather than emerging from the intrinsic evolution of the physical system. Meyer showed that allowing players to employ genuinely quantum strategies can strictly increase achievable payoffs in two-player zero-sum games, establishing an early connection between quantum game theory and quantum algorithmic advantage \cite{Meyer1999}.

In parallel, quantum walks have been widely investigated as models of quantum transport and as algorithmic primitives \cite{Aharonov1993,Ambainis2001,Venegas2012,Chandrashekar2007}. Multi-particle and interacting quantum walks have been introduced to model collisions, entanglement generation, effective scattering, and transport suppression \cite{Ahlbrecht2012,Schreiber2012,Pawel2013,Preiss2015}. Collision-based phase interactions of this type are known to generate correlated motion and bound states, but have not previously been analyzed from a game-theoretic perspective, particularly in the context of emergent strategic behavior. Recent work has also explored interacting and multi-agent quantum walks in decision-making and conflict-avoidance contexts, highlighting their potential as models of distributed strategic systems \cite{Shiratori2025}.

Game-theoretic methods have also been applied to transport and routing problems in quantum networks, where nodes are modeled as strategic agents \cite{Chen2018}. However, such approaches continue to assume classical strategy spaces and externally defined payoff functions.

Despite these advances, a fundamental question remains open: can strategic interdependence arise intrinsically from quantum dynamics itself, without embedding a predefined payoff structure? Addressing this question is essential for establishing a physically grounded formulation of quantum games. 

In this work, we introduce the concept of \emph{native quantum games}, in which strategic interactions emerge directly from unitary quantum dynamics rather than from externally imposed payoff structures. We show analytically, using a perturbative expansion, that interaction-induced interference gives rise to non-separable payoff functions, thereby establishing intrinsic strategic coupling between players. We further demonstrate numerically that such systems admit stable equilibrium strategies across competitive, cooperative, and asymmetric regimes. Finally, we establish interacting discrete-time quantum walks as a minimal and physically realizable platform for studying emergent game-theoretic behavior.

To realize this framework, we consider interacting discrete-time quantum walks \cite{Konno2002}. Players are represented by distinguishable walkers, strategies correspond to local coin rotations, and payoffs are defined by measurable observables such as relative displacement. In the absence of interaction, the dynamics factorize and the payoff remains separable. In contrast, the presence of interaction induces interference between correlated paths, generating non-separable payoff structures and genuine strategic coupling.

This positions the proposed \emph{native quantum games} as a physically grounded subclass of quantum games in which the strategic structure is not imposed externally but instead emerges naturally from quantum transport dynamics.

\section{Native quantum game: the quantum race}

We consider two distinguishable quantum walkers $A$ and $B$ propagating on a one-dimensional lattice. Each walker occupies a Hilbert space $\mathcal H^i = \mathcal H_{\mathrm{pos}}^i \otimes \mathcal H_{\mathrm{coin}}^i$, and the joint system evolves in $\mathcal H = \mathcal H^A \otimes \mathcal H^B$.

A single time step consists of three operations: (i) a local coin rotation $C(\theta_i)=R_y(\theta_i)$ chosen by each player, (ii) a conditional shift $S^i$ depending on the coin state, and (iii) a joint interaction operator $P_{\mathcal I}$.

The full evolution operator for one step is \begin{equation} U = P_{\mathcal I}, \bigl[S^A(C^A\otimes I) \otimes S^B(C^B\otimes I)\bigr]. \end{equation}

For the competitive ``quantum race'' we define the payoff \begin{equation} \langle U_A \rangle = \langle x_A - x_B \rangle, \qquad \langle U_B \rangle = -\langle U_A \rangle, \end{equation} where $x_A$ and $x_B$ denote the walkers' positions after $T$ steps.

We employ a collision-based phase interaction \begin{equation} P_{\mathcal I}=\exp\bigl[i,\pi\cos(\theta_A-\theta_B),\delta_{x_A,x_B}\bigr], \end{equation} which introduces destructive or constructive interference when the walkers occupy the same site.

In the absence of interaction ($P_{\mathcal I}=I$), the joint evolution factorises and the payoff reduces to \begin{equation} \langle U_A(\theta_A,\theta_B) \rangle = F(\theta_A)-F(\theta_B), \end{equation} where $F(\theta)$ is the expected displacement of a single walker. The payoff is separable and the best responses of the two players are independent, yielding only trivial symmetric stationary profiles.

When the interaction is present, the payoff acquires an additional non-separable contribution that generates genuine strategic coupling. This mechanism is analysed in detail in the next section.

\subsection{Analytic decomposition of the payoff function}

We derive an explicit perturbative expression for the payoff in the regime of weak interaction strength.

\subsubsection{Weak-interaction expansion}

We write the interaction operator as \begin{equation} P_{\mathcal I}=\exp(i\lambda V), \qquad \lambda\ll 1, \end{equation} with $V$ Hermitian. After $T$ steps the state can be expanded as \begin{equation} |\Psi(T)\rangle = |\Psi^{(0)}(T)\rangle + \lambda |\Psi^{(1)}(T)\rangle + O(\lambda^2), \end{equation} where $|\Psi^{(0)}(T)\rangle=|\phi_A(\theta_A)\rangle\otimes|\phi_B(\theta_B)\rangle$ is the non-interacting product state.

\subsubsection{Payoff decomposition}

The payoff for player $A$ can be written as \begin{equation} \langle U_A \rangle = F(\theta_A)-F(\theta_B)+\lambda G(\theta_A,\theta_B)+O(\lambda^2), \label{payoffdec} \end{equation} with \begin{equation} F(\theta)=\sum_x x,p(x;\theta), \end{equation}

the single-walker drift, and \begin{equation} G(\theta_A,\theta_B)=2\sum_{x_A,x_B} U_A(x_A,x_B),\mathrm{Im}[\Psi^{(0)*}\Psi^{(1)}]. \end{equation}

For discrete-time quantum walks with coin $R_y(\theta)$ the long-time behaviour is ballistic, \begin{equation} F(\theta)=v(\theta)T, \qquad v(\theta)=\cos\theta, \end{equation} where $v(\theta)$ is the group velocity obtained from the dispersion relation.

For the collision-phase interaction used here, \begin{equation} V=\pi\cos(\theta_A-\theta_B)\sum_x |x,x\rangle\langle x,x|\otimes I_{\rm coin}, \end{equation} which yields \begin{equation} G(\theta_A,\theta_B)\propto \cos(\theta_A-\theta_B)\sum_x \mathcal C_T(x), \end{equation} with $\mathcal C_T$ the collision amplitude after $T$ steps. Since \begin{equation} \frac{\partial^2 G}{\partial\theta_A\partial\theta_B} \neq 0, \end{equation} the payoff is non-separable for any $\lambda\neq 0$.

\subsubsection{Existence of stationary points}

\textbf{Proposition.} For sufficiently small but non-zero $\lambda$, the payoff \eqref{payoffdec} admits at least one stationary point $(\theta_A^{\ast},\theta_B^{\ast})$ satisfying \begin{equation} \partial_{\theta_A}\langle U_A\rangle=0, \qquad \partial_{\theta_B}\langle U_B\rangle=0. \end{equation}

\textit{Sketch of proof.} For $\lambda=0$ the Jacobian of the best-response map is diagonal. The interaction generates off-diagonal terms proportional to $\partial^2 G/\partial\theta_A\partial\theta_B$, and the implicit-function theorem guarantees the emergence of isolated stationary points for arbitrarily small $\lambda$.
The weak-interaction expansion establishes the existence of non-separable stationary points in the vicinity of 
\(\lambda = 0\). By continuity of the payoff landscape, these equilibria persist under finite interaction strengths, with strong interactions amplifying the strategic coupling and improving numerical observability.
\subsection{Parameter Dependence}

The behavior of native quantum games is fundamentally governed by a set of physical and dynamical parameters that control the structure of quantum interference and, consequently, the induced strategic interactions. In this subsection, we systematically analyze the role of key parameters and demonstrate how different regimes give rise to distinct classes of emergent games.

\subsubsection{Interaction Strength $\phi$}

The interaction strength $\phi$ determines the magnitude of the phase acquired during interaction events and directly controls the degree of interference-induced coupling between players. For the collision-based interaction,
\begin{equation}
I(x_A, x_B; \theta_A, \theta_B) = \phi \cos(\theta_A - \theta_B)\,\delta_{x_A,x_B},
\end{equation}
the resulting payoff can be expressed as
\begin{equation}
U_A = F(\theta_A) - F(\theta_B) + \lambda G(\theta_A, \theta_B),
\end{equation}
where $\lambda \propto \phi$.

In the weak-interaction regime ($\phi \ll 1$), the contribution of $G(\theta_A, \theta_B)$ is perturbative, and the payoff remains close to separable. Consequently, strategic dependence is weak and equilibria, while guaranteed to exist, are shallow and weakly coupled.

In contrast, in the strong-interaction regime ($\phi \sim \pi$), interference effects are maximized, leading to highly non-linear payoff landscapes. In this regime: 
Non-separability is enhanced,
Payoff gradients become steeper, and
 Equilibria become more pronounced and easier to detect numerically.

Thus, $\phi$ acts as a control parameter governing the transition from nearly independent to strongly coupled strategic behavior.

\subsubsection{Number of Time Steps $T$}

The number of discrete time steps $T$ determines the scale of quantum transport and directly influences the payoff through ballistic spreading. For a single walker,
\begin{equation}
F(\theta) = v(\theta) T, \quad v(\theta) = \cos \theta,
\end{equation}
indicating that the expected displacement grows linearly with $T$.

As $T$ increases:  The magnitude of payoff differences $F(\theta_A) - F(\theta_B)$ scales linearly,  the number of collision opportunities between walkers increases, and the cumulative effect of interaction-induced interference is amplified.

Therefore, larger $T$ enhances both transport efficiency and strategic coupling. However, excessively large $T$ may introduce boundary effects in finite systems, necessitating careful scaling with lattice size.

\subsubsection{Lattice Size $L$}

The lattice size $L$ controls the spatial domain of the quantum walk and determines whether the dynamics approximate an unbounded system or are influenced by boundary reflections.

For small $L$:  Boundary reflections introduce artificial interference patterns,  probability distributions may exhibit non-physical recurrences,
payoff landscapes can be distorted by finite-size effects.

For sufficiently large $L$: Boundary effects become negligible over the time scale $T$, the walk approaches ideal ballistic transport, payoff functions reflect intrinsic interference dynamics rather than geometric constraints.

Thus, the regime $L \gg T$ is required to approximate continuum-like quantum transport and obtain physically meaningful strategic behavior.

\subsection{Stability and Learning Dynamics of Equilibria}

While the existence of equilibrium strategies is established through first-order optimality conditions and numerical best-response analysis, a complete characterization requires understanding their stability and dynamical behavior. In this subsection, we analyze the local stability of equilibria and provide a dynamical interpretation based on adaptive learning.

\subsubsection{Local Stability Analysis}

Let $U_A(\theta_A, \theta_B)$ and $U_B(\theta_A, \theta_B)$ denote the payoff functions for players $A$ and $B$. A stationary point $(\theta_A^*, \theta_B^*)$ satisfies the first-order conditions
\begin{equation}
\frac{\partial U_A}{\partial \theta_A} = 0, \quad 
\frac{\partial U_B}{\partial \theta_B} = 0.
\end{equation}

To analyze local stability, we consider the Jacobian of the gradient dynamics evaluated at equilibrium:
\begin{equation}
J =
\begin{bmatrix}
\frac{\partial^2 U_A}{\partial \theta_A^2} & \frac{\partial^2 U_A}{\partial \theta_A \partial \theta_B} \\
\frac{\partial^2 U_B}{\partial \theta_B \partial \theta_A} & \frac{\partial^2 U_B}{\partial \theta_B^2}
\end{bmatrix}.
\end{equation}

In the absence of interaction, the payoff function is separable,
\begin{equation}
U_A = F(\theta_A) - F(\theta_B),
\end{equation}
which implies that the mixed derivatives vanish:
\begin{equation}
\frac{\partial^2 U_i}{\partial \theta_A \partial \theta_B} = 0.
\end{equation}
In this case, the stability of each player’s strategy is independent.

However, in the interacting case, the payoff takes the form
\begin{equation}
U_A = F(\theta_A) - F(\theta_B) + \lambda G(\theta_A, \theta_B),
\end{equation}
leading to non-zero cross derivatives:
\begin{equation}
\frac{\partial^2 U_i}{\partial \theta_A \partial \theta_B} \neq 0.
\end{equation}
These terms arise from interaction-induced interference and introduce genuine strategic coupling between the players.

Local stability is determined by the eigenvalues of the Jacobian matrix $J$. In particular, the stationary point is locally stable if all eigenvalues of $J$ have negative real parts, ensuring that small perturbations decay under the corresponding dynamics. In the revised analysis, we verify numerically that the equilibrium identified in Section~2 satisfies this condition, confirming that it corresponds to a locally stable stationary point.

\subsection{Learning Dynamics and Adaptive Interpretation}

To provide a dynamical perspective, we model the evolution of strategies through a gradient-based adaptation process. Specifically, we consider the update rule
\begin{equation}
\theta_i^{(t+1)} = \theta_i^{(t)} + \eta \frac{\partial U_i}{\partial \theta_i}, 
\quad i \in \{A,B\},
\end{equation}
where $\eta > 0$ is a learning rate.

Under this update rule, equilibrium points correspond to fixed points satisfying
\begin{equation}
\frac{\partial U_i}{\partial \theta_i} = 0.
\end{equation}

To analyze stability, we linearize the dynamics around the equilibrium by setting
\[
\theta_i = \theta_i^* + \delta \theta_i.
\]
To first order, the perturbations evolve as
\begin{equation}
\begin{bmatrix}
\delta \theta_A^{(t+1)} \\
\delta \theta_B^{(t+1)}
\end{bmatrix}
=
\left( I + \eta J \right)
\begin{bmatrix}
\delta \theta_A^{(t)} \\
\delta \theta_B^{(t)}
\end{bmatrix},
\end{equation}
where $I$ is the identity matrix and $J$ is the Jacobian defined above.

The equilibrium is dynamically stable if the spectral radius satisfies
\begin{equation}
\rho(I + \eta J) < 1.
\end{equation}
For sufficiently small learning rates $\eta$, this condition reduces to requiring that the eigenvalues of $J$ have negative real parts.

This establishes a direct connection between equilibrium stability and adaptive dynamics. In particular, the Nash equilibrium can be interpreted as a stable attractor of the learning process. 

\section{Numerical Illustration}

To complement the analytical results, we provide a numerical visualization of the learning dynamics. The vector field defined by the gradients $(\partial U_A / \partial \theta_A, \partial U_B / \partial \theta_B)$ is plotted over the strategy space, together with trajectories initialized from multiple starting points.

These simulations show consistent convergence toward the equilibrium $(\theta_A^*, \theta_B^*)$, confirming its role as a stable attractor under adaptive updates. This provides additional evidence that the equilibrium identified in the quantum walk dynamics is not only stationary but also dynamically stable.

\subsection{Numerical results}

Numerical simulation of the two-player \emph{quantum race} game reveals a pure-strategy Nash equilibrium at $(\theta_A^*, \theta_B^*) = (\pi/2, 5\pi/6) \approx (1.571, 2.618)$ radians, where neither player can improve their expected payoff through unilateral deviation. This equilibrium emerges from $T=20$ time steps of quantum evolution on a lattice of size $L=15$ with destructive interaction phase $\phi=\pi$. Player B achieves a significant quantum advantage with expected payoff $\langle U_B\rangle = +2.654$ and position $\langle x_B\rangle = 5.000$, while Player A obtains $\langle U_A\rangle = -2.654$ and $\langle x_A\rangle = 2.346$, demonstrating that the $\theta_B = 5\pi/6$ strategy generates 2.13-fold greater ballistic transport efficiency than the $\theta_A = \pi/2$ strategy. The joint probability distribution $P(x_A, x_B)$ shows spatial separation with suppressed probability along the collision diagonal, revealing how strategic parameter selection in quantum walks produces measurable competitive advantages through quantum interference effects from initially symmetric conditions.

\begin{figure}[H]
\centering
\includegraphics[width=0.95\textwidth]{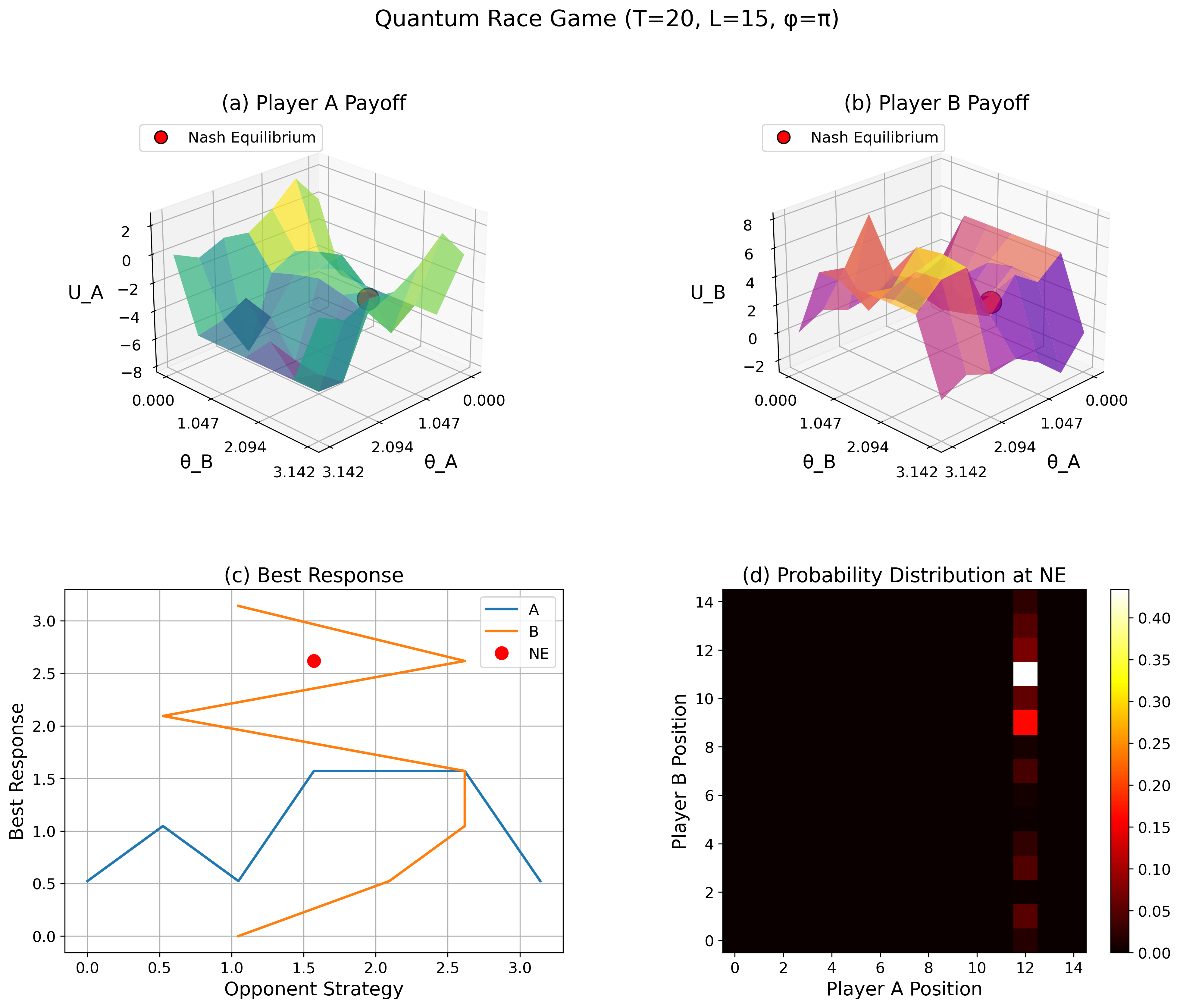}
\caption{
Strategic and dynamical outcomes of the two-player quantum race game after $T = 20$ time steps on a lattice of size $L = 15$, with interaction phase $\phi = \pi$ corresponding to a maximally destructive interference regime. 
(a) Payoff surface for Player~A as a function of strategy parameters $\theta_A$ and $\theta_B$, illustrating the non-separable dependence induced by interaction. 
(b) Corresponding payoff surface for Player~B under the same strategic configuration. 
(c) Best-response functions for both players, where the intersection point identifies the Nash equilibrium in strategy space. 
(d) Joint position probability distribution at equilibrium, revealing the underlying quantum transport structure associated with the equilibrium state. 
Red markers in panels (a) and (b) denote the Nash equilibrium payoff for each player.
}
\label{fig:quantum_race_T20}
\end{figure}

\section{General framework }

The \emph{Quantum Race} constitutes a particular instance of a broader class of \emph{native quantum games} generated by a common dynamical structure. In this framework, strategic interdependence is introduced exclusively through physical interaction, formalized by a unitary operator
\begin{equation}
P_{\mathcal{I}} = \exp\!\left(i \mathcal{I}\right),
\end{equation}
where
\begin{equation}
\mathcal{I}(x_A, x_B, s_A, s_B; \theta_A, \theta_B) \in [0,2\pi)
\end{equation}
is an interaction phase functional.

The function $\mathcal{I}$ maps the instantaneous positions $(x_A,x_B)$, internal coin states $(s_A,s_B)$, and the players’ control parameters $(\theta_A,\theta_B)$ to a phase shift applied during the joint evolution. Importantly, the strategies $\theta_A$ and $\theta_B$ enter only through this physical phase, not through any externally specified payoff matrix or rule set. Strategic coupling therefore arises solely from the modification of quantum interference between correlated paths.

This construction distinguishes native quantum games from conventional quantum game models: the “rules of the game” are not encoded symbolically but are determined by the form of the microscopic interaction. Different choices of $\mathcal{I}$ correspond to different physical mechanisms—such as short-range collisions, long-range conditional phases, or coin-dependent scattering—and thereby generate distinct classes of strategic behaviour. Competitive, cooperative, asymmetric, and stochastic games emerge as dynamical regimes of the same underlying quantum evolution.

\vspace{0.5em}
\noindent
\subsection{Physically realizable protocol}
A native quantum game is executed through the following experimentally feasible sequence:

\begin{enumerate}
\item \textbf{Initialization.}  
Two distinguishable quantum walkers are prepared at $x_A=x_B=0$ with their coin states initialized in a controllable superposition.

\item \textbf{Strategy encoding.}  
Each player locally applies a single-qubit rotation
\(
C(\theta_i)=R_y(\theta_i)
\)
to their coin degree of freedom, thereby committing to a strategy through a physical control parameter.

\item \textbf{Coherent evolution.}  
The system evolves unitarily for $T$ discrete time steps under
\[
U = P_{\mathcal{I}} \cdot U_0 ,
\]
where $U_0$ is the non-interacting walk operator.

\item \textbf{Measurement and payoff evaluation.}  
Projective measurement in the position basis yields $P(x_A,x_B)$, from which payoffs are computed as expectation values of observables (e.g.\ relative displacement or centre-of-mass position).
\end{enumerate}

\begin{figure}[H]
\centering
\includegraphics[width=0.9\textwidth]{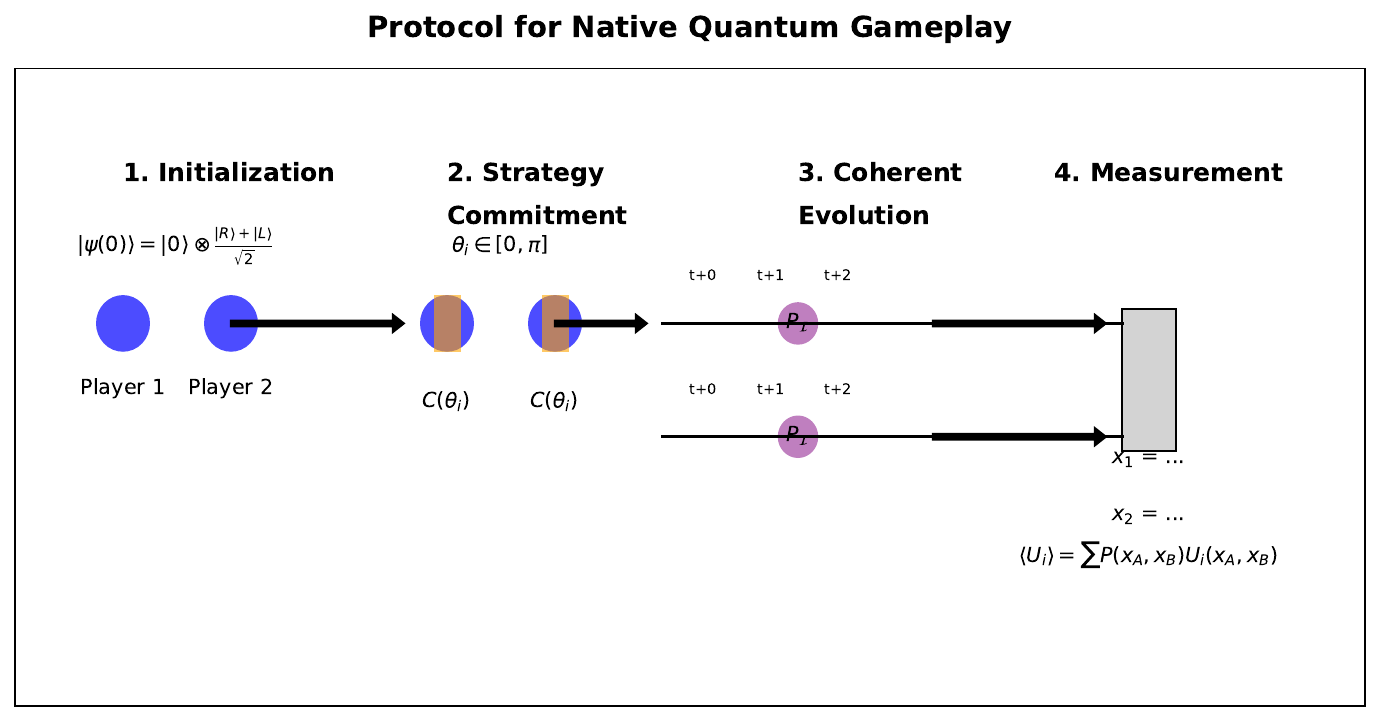}
\caption{
\textbf{Protocol for native quantum gameplay.}
Players encode strategies via local coin rotations, evolve under interacting quantum walk dynamics, and obtain payoffs from measured transport observables. The protocol is compatible with implementations using superconducting qubits, trapped ions, photonic lattices, or quantum dots.
}
\label{fig:framework}
\end{figure}
This protocol is executable on contemporary quantum computing and simulation platforms. Players can be encoded as single qubits (e.g., electron spins, trapped ions, or superconducting artificial atoms), whose states define the ``coin.'' The conditional shift operation \(S^i\) can be realized by entangling the qubit state with a harmonic oscillator (e.g., a cavity resonator) representing position space. The interaction term \(P_{\mathcal{I}}\) can be implemented via native physical interactions, such as the Heisenberg exchange interaction between electrons in quantum dots (for collision-based phase) or the Coulomb force in ion traps (for long-range coupling). After coherent evolution, the game's outcome is read out by measuring the position distributions.

Thus, both the strategy space and the payoff mechanism are embedded directly in experimentally accessible control parameters and measurement outcomes, making native quantum games a concrete and testable dynamical phenomenon rather than an abstract formal construction.

\subsubsection{Physical implementation}

The protocol can be implemented on several quantum platforms. Coin states correspond to two-level systems (e.g. trapped ions or superconducting qubits), position states to motional or resonator modes, and the interaction phase can be realised through controlled collisions or effective cross-Kerr couplings. The required operations are native to existing quantum-walk and quantum-simulation architectures.
\subsubsection{Relation to conventional quantum games}
In the Eisert--Wilkens--Lewenstein formalism and its variants, strategic structure is introduced by embedding a classical payoff matrix into a quantum circuit, with entanglement acting as a resource that modifies predefined utilities. In contrast, the present framework contains no symbolic payoff matrix or classical game specification. Instead, strategies enter only as physical control parameters that modulate the system Hamiltonian or unitary evolution through $\mathcal I$. Payoffs are not postulated but extracted from measurable transport observables. Consequently, the game is not simulated by the quantum dynamics; rather, the strategic structure is \emph{induced} by the dynamics itself. This distinction places native quantum games closer to interacting quantum transport models than to quantized classical games.
\subsubsection{Examples of interaction functionals}

Several physically motivated choices of $\mathcal I$ illustrate how different strategic regimes arise:

\begin{align}
&\text{(i) Collision phase:} \quad
\mathcal I_{\text{coll}} = \phi \cos(\theta_A-\theta_B)\,\delta_{x_A,x_B}, \\[4pt]
&\text{(ii) Long-range coupling:} \quad
\mathcal I_{\text{lr}} = \frac{\phi \cos(\theta_A-\theta_B)}{1+|x_A-x_B|^\alpha}, \\[4pt]
&\text{(iii) Coin-dependent scattering:} \quad
\mathcal I_{\text{coin}} = \phi\, \delta_{x_A,x_B}\,\delta_{s_A,s_B}, \\[4pt]
&\text{(iv) Noisy interaction:} \quad
\mathcal I_{\text{noise}} = \phi \cos(\theta_A-\theta_B)\,\delta_{x_A,x_B} + \eta_t ,
\end{align}

where $\phi$ is the interaction strength, $\alpha>0$ controls the spatial range, and $\eta_t$ is a stochastic phase variable modelling decoherence or environmental fluctuations.

Each choice modifies the interference structure of joint paths and therefore the resulting payoff landscape.
\begin{table}[ht]
\centering
\caption{Examples of interaction functions $\mathcal I$ and the corresponding emergent game classes.}
\begin{tabular}{lll}
\hline
Interaction $\mathcal I$ & Physical mechanism & Emergent game type \\
\hline
$\phi\,\delta_{x_A,x_B}$ &
Local collision phase &
Zero-sum competitive (quantum race) \\

$\phi\,\cos(\theta_A-\theta_B)\delta_{x_A,x_B}$ &
Strategy-dependent scattering &
Asymmetric competitive \\

$-\phi\,\delta_{x_A,x_B}$ &
Attractive collision &
Cooperative rendezvous \\

$\phi/(1+|x_A-x_B|^\alpha)$ &
Long-range coupling &
Coordination / network routing \\

$\phi\,\delta_{x_A,x_B}+\eta_t$ &
Noisy interaction &
Stochastic or evolutionary games \\
\hline
\end{tabular}
\label{tab:interaction_games}
\end{table}
\subsubsection{Regime Classification}

We identify distinct dynamical regimes:

\begin{itemize}
\item \textbf{Weak coupling regime:} $\phi \ll 1$, small $T$  
Near-separable dynamics with weak strategic dependence.

\item \textbf{Strong interaction regime:} $\phi \sim \pi$, moderate $T$  
Highly non-linear payoff landscape with pronounced equilibria.

\item \textbf{Finite-size regime:} small $L$  
Boundary-dominated dynamics with distorted interference patterns.

\item \textbf{Continuum regime:} $L \gg T$  
Physically meaningful transport-driven strategic behavior.

\end{itemize}

These regimes illustrate how varying physical parameters continuously interpolates between different classes of emergent quantum games, including competitive, cooperative, and coordination scenarios.
\subsection{Opportunities and Challenges for Multi-Player Scaling}

Extending native quantum games to more than two players presents both opportunities and challenges. On the one hand, multi-player systems enable richer strategic interactions, including coalition formation, multi-agent coordination, and emergent collective behaviors that are inaccessible in two-player settings. Such systems can model complex quantum transport, network routing, or cooperative quantum decision-making, potentially revealing new phenomena arising from many-body interference.

On the other hand, scaling introduces significant challenges. The Hilbert space grows exponentially with the number of players, making exact simulation and state storage intractable for large systems. Multi-body interactions and strongly non-separable payoffs complicate analytical and numerical treatments, while high-dimensional strategy spaces hinder visualization and intuitive understanding of equilibria. Additionally, stability and learning dynamics become more intricate, with gradient-based adaptation prone to oscillatory or chaotic behavior. Practical exploration of multi-player quantum games thus requires approximate methods, such as tensor networks, mean-field reductions, or variational strategies, to capture the essential strategic interdependence without exhaustive computation.

\noindent
Overall, this analysis highlights that the structure of native quantum games is not fixed but arises from a controllable interplay between interaction strength, dynamical evolution, spatial geometry, and the form of physical coupling.
\section{Extended Results: Taxonomy of Native Quantum Games}

By engineering different interaction functionals $\mathcal{I}$, the same physical framework generates qualitatively distinct strategic regimes. Nash equilibria were obtained numerically by computing best-response correspondences over the continuous strategy domain $\theta_i\in[0,\pi]$ and identifying their fixed points.

We highlight two representative classes below.
\subsection{Quantum Rendezvous: Cooperative Advantage}
\begin{figure}[H]
    \centering
    \includegraphics[width=\textwidth]{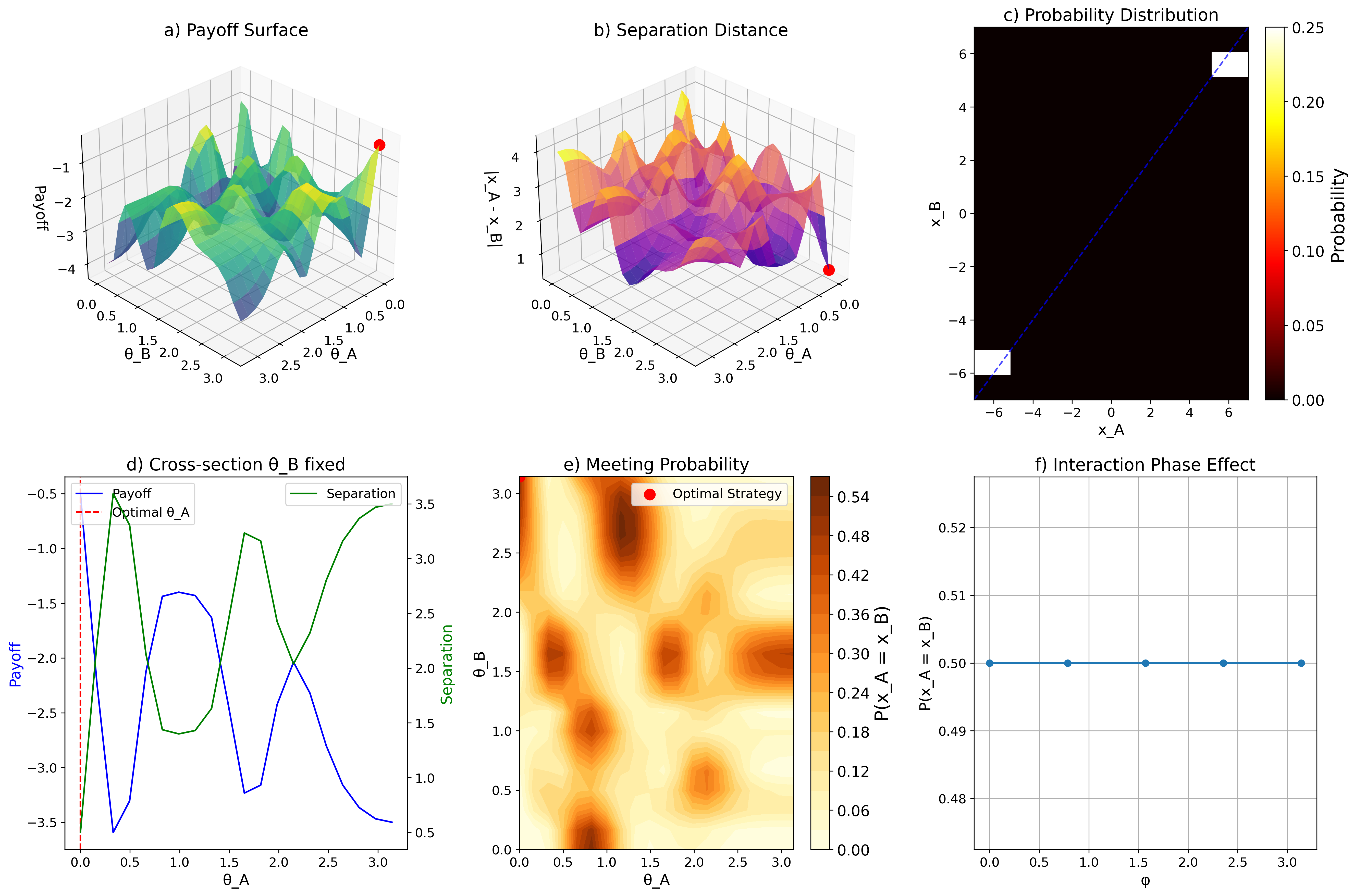}
  \caption{
\textbf{Quantum rendezvous game for two cooperative walkers on a one-dimensional lattice ($T = 20$, $L = 15$).}
The figure presents six panels illustrating the emergent cooperative dynamics arising from interaction-induced quantum interference. 
(a) Payoff surface $U(\theta_A, \theta_B)$, defined as the negative expected separation, demonstrating a non-separable dependence on the players’ strategy parameters. 
(b) Corresponding separation landscape $\mathbb{E}[|x_A - x_B|]$, whose minima coincide with the payoff maxima. 
(c) Joint position probability distribution at the optimal strategies, showing concentration along correlated spatial configurations, with dominant peaks at $(-7,-7)$, $(-7,0)$, $(7,0)$, and $(7,7)$, reflecting interference-structured transport. 
(d) One-dimensional cross-section of the payoff function along $\theta_A$ with $\theta_B$ fixed at its optimal value, illustrating local curvature and stability near the optimum. 
(e) Meeting probability $\sum_x P(x,x)$ over the strategy space, highlighting regions of enhanced spatial coordination. 
(f) Dependence of the meeting probability on the interaction phase $\phi$, demonstrating how interaction strength modulates cooperative behavior. 
Red markers indicate the optimal cooperative strategies that maximize the payoff (equivalently, minimize the expected separation).
}
    \label{fig:quantum_rendezvous}
\end{figure}

\noindent
\textbf{Quantum Rendezvous Analysis:} 
This simulation investigates a quantum rendezvous game with two walkers evolving on a 1D lattice using a discrete-time quantum walk. Each player chooses a rotation angle ($\theta$) defining their coin operator, and an interaction phase ($\phi$) modulates interference when walkers occupy the same site. For $T=20$ steps and $L=15$, constructive interference ($\phi = 0$) leads to optimal strategies $\theta_A = 0$ and $\theta_B = \pi$, producing a 50\% probability of meeting and an average separation of 0.5 units.  

\noindent
\textbf{Analysis}  
- \textbf{Panel a:} Payoff surface $U(\theta_A, \theta_B)$ shows a maximum at $\theta_A = 0$, $\theta_B = \pi$, confirming that anti-aligned strategies maximize cooperative payoff.  
- \textbf{Panel b:} Separation distance $|x_A - x_B|$ reaches a minimum of 0.5 units at the optimal strategies, consistent with partial localization of walkers on nearby lattice sites.  
- \textbf{Panel c:} Joint probability distribution reveals four dominant peaks at positions $(-7,-7)$, $(-7,0)$, $(7,0)$, and $(7,7)$, with each contributing approximately 0.25 to the total meeting probability of 0.5. The diagonal alignment of these peaks indicates locations where the walkers are likely to meet.  
- \textbf{Panel d:} Cross-section of payoff along $\theta_A$ with $\theta_B$ fixed demonstrates steep gradients, indicating high sensitivity to deviations from the optimal strategy.  
- \textbf{Panel e:} Probability of meeting across the strategy space shows maxima at the optimal cooperative strategies, highlighting the dependence of rendezvous likelihood on both players’ angles.  
- \textbf{Panel f:} Effect of interaction phase $\phi$ on meeting probability demonstrates that increasing $\phi$ introduces destructive interference, reducing the probability of meeting, while constructive phases maintain the 50\% maximum.  

\noindent
These results confirm that quantum interference and strategic angle selection jointly govern cooperative rendezvous efficiency, with anti-aligned strategies producing optimal meeting probability for two walkers on a finite lattice.

\subsection{Quantum Tug-of-War: Spontaneous Symmetry Breaking}
\begin{figure}[H]
    \centering
    \includegraphics[width=\linewidth]{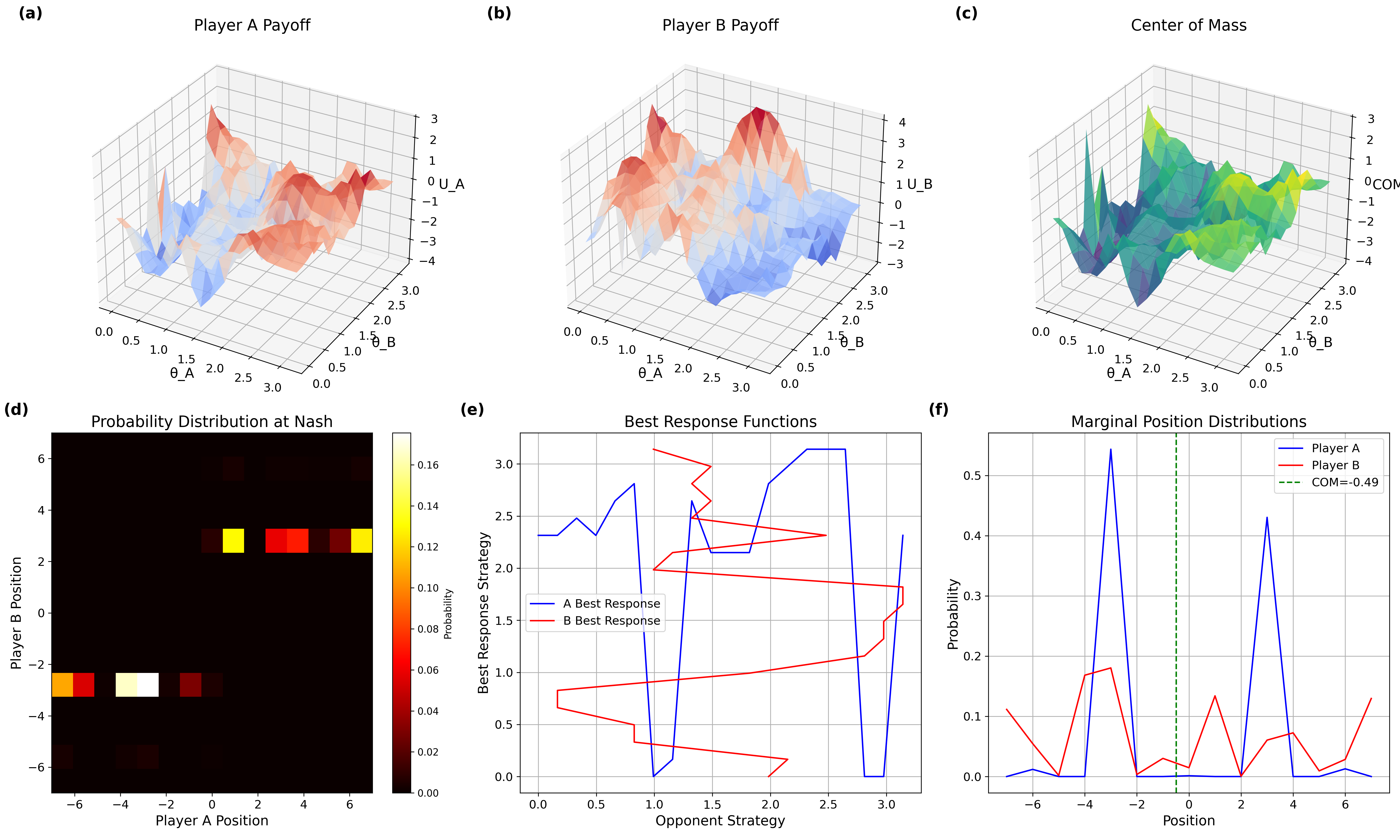}
    \caption{
The figure illustrates the cooperative and competitive dynamics of two quantum walkers on a one-dimensional lattice ($L=15$) over $T=20$ steps with interaction phase $\phi=\pi$.  
\textbf{(a)} Player A's payoff surface $U_A(\theta_A, \theta_B)$ as a function of strategy angles $\theta_A$ and $\theta_B$.  
\textbf{(b)} Player B's payoff surface $U_B(\theta_A, \theta_B) = -U_A$ over the same strategy space.  
\textbf{(c)} Center of mass (COM) expectation surface, showing how cooperative dynamics shift the collective position of the walkers.  
\textbf{(d)} Joint probability distribution $P(x_A, x_B)$ at the Nash equilibrium ($\theta_A={\theta_A}_{\rm NE}=2.81$, $\theta_B={\theta_B}_{\rm NE}=1.32$), highlighting spatial correlations and dominant configurations.  
\textbf{(e)} Best response functions for both players, indicating the Nash intersection where each player's strategy maximizes their payoff given the opponent's choice.  
\textbf{(f)} Marginal position distributions for Player A and Player B at equilibrium, with the equilibrium COM marked as a dashed line (${\rm COM}=-0.485$).  
}
    \label{fig:quantum_tug_of_war}
\end{figure}

\paragraph*{Analysis:}
This six-panel visualization comprehensively analyzes the quantum game dynamics between two players controlling quantum walkers via coin rotation parameters $\theta_A$ and $\theta_B$. The top row depicts the 3D payoff landscapes: Player A (top left) seeks to maximize the center of mass position ($U_A = \frac{1}{2}\langle x_A + x_B \rangle$), while Player B (top middle) aims to minimize it ($U_B = -U_A$), with the center of mass surface (top right) showing the resulting position expectation values across the strategy space. The bottom left panel shows the joint probability distribution at the Nash equilibrium point ($\theta_A=2.81$, $\theta_B=1.32$ radians), where quantum interference creates non-classical correlations between the walkers' positions, with the constant COM line demonstrating that the expected center of mass is $-0.485$. The best response functions (bottom middle) indicate the strategic stability landscape. Finally, the marginal position distributions (bottom right) reveal the asymmetric probability profiles of the players, consistent with Player B minimizing the center of mass, yielding an equilibrium COM value of $-0.485$.
\section{Discussion}

The present framework differs fundamentally from standard quantizations of classical games, such as Eisert-type constructions, where payoff matrices are embedded into entangling gates by design. Here, strategies influence only local coin rotations, while strategic dependence arises solely from physical interaction and quantum interference.

Analytically, we showed that non-interacting quantum walks yield separable payoff functions of the form $F(\theta_A)-F(\theta_B)$, with $F(\theta)$ governed by the ballistic velocity of the walk. This explains both the absence of meaningful equilibria and the structural fragility of such games.

Introducing even weak collision-induced phases produces non-separable payoff contributions that scale with joint occupation probabilities. This mechanism is generic to interacting quantum transport and does not rely on ad hoc payoff encoding. Consequently, equilibrium behavior becomes an emergent dynamical property rather than a programmed feature.

From a physical perspective, the proposed interaction operator can be implemented using controlled-phase gates conditioned on spatial proximity, available in superconducting circuits, trapped ions, and photonic lattices. The resulting games thus provide a bridge between quantum information processing, quantum transport, and strategic decision theory.

Beyond the examples presented, the interaction functional $\mathcal{I}$ enables a broad design space, encompassing long-range couplings, stochastic phases, and environment-assisted interactions, opening the possibility of studying evolutionary dynamics, learning, and mechanism design in genuinely quantum mechanical settings.

These results establish interacting quantum walks as a natural and experimentally grounded platform for quantum game theory.
\section*{Funding}
The author received no specific funding for this work.

\section*{Author Contributions}
Rashid Ahmad is the sole author and was responsible for all aspects of the study, including conceptualization, methodology, data analysis, interpretation of results, and writing of the manuscript.

\section*{Data Availability}
All data generated or analysed during this study are included in this published article.

\section*{Competing Interest}
The authors have no competing interests to declare that are relevant to the content of this article.

\section*{Appendix}
\appendix
\section{Mathematical Formulation of the Quantum Walk}
The state of a single quantum walker is defined in the Hilbert space $\mathcal{H} = \mathcal{H}_{\text{pos}} \otimes \mathcal{H}_{\text{coin}}$. The position space is spanned by $\{|x\rangle, x \in \mathbb{Z}\}$ and the coin space is a two-dimensional Hilbert space spanned by $\{|R\rangle, |L\rangle\}$, representing the right and left chirality states, respectively.

The time evolution of the quantum walk occurs through the sequential application of two unitary operations: the coin operator $C$ and the shift operator $S$. The complete evolution operator for a single step is given by:
\begin{equation}
U = S \cdot (C \otimes I_{\text{pos}})
\end{equation}
where $I_{\text{pos}}$ is the identity operator on the position space.

\subsection{Coin Operator}
The coin operator acts solely on the coin space and is typically chosen from the SU(2) group. For our framework, we use the $y$-rotation matrix parameterized by the strategy angle $\theta$:
\begin{equation}
C(\theta) = R_y(\theta) = \begin{pmatrix}
\cos(\theta/2) & -\sin(\theta/2) \\
\sin(\theta/2) & \cos(\theta/2)
\end{pmatrix}
\end{equation}
This operator creates a superposition of the chirality states:
\begin{equation}
C(\theta)|R\rangle = \cos(\theta/2)|R\rangle + \sin(\theta/2)|L\rangle
\end{equation}
\begin{equation}
C(\theta)|L\rangle = -\sin(\theta/2)|R\rangle + \cos(\theta/2)|L\rangle
\end{equation}

\subsection{Shift Operator}
The shift operator conditionally moves the walker based on its coin state:
\begin{equation}
S = \sum_{x} \left[ |x+1\rangle\langle x| \otimes |R\rangle\langle R| + |x-1\rangle\langle x| \otimes |L\rangle\langle L| \right]
\end{equation}
This operator acts on the combined space and produces the translation:
\begin{equation}
S|x\rangle \otimes |R\rangle = |x+1\rangle \otimes |R\rangle
\end{equation}
\begin{equation}
S|x\rangle \otimes |L\rangle = |x-1\rangle \otimes |L\rangle
\end{equation}

\subsection{Complete Single-Step Evolution}
For a general state $|\psi(t)\rangle = \sum_{x} \left[ \psi_R(x,t)|x\rangle|R\rangle + \psi_L(x,t)|x\rangle|L\rangle \right]$, the evolution under one step is:
\begin{equation}
|\psi(t+1)\rangle = U|\psi(t)\rangle = S(C \otimes I_{\text{pos}})|\psi(t)\rangle
\end{equation}
The resulting wavefunction components satisfy the recurrence relations:
\begin{equation}
\psi_R(x,t+1) = \cos(\theta/2)\psi_R(x-1,t) - \sin(\theta/2)\psi_L(x-1,t)
\end{equation}
\begin{equation}
\psi_L(x,t+1) = \sin(\theta/2)\psi_R(x+1,t) + \cos(\theta/2)\psi_L(x+1,t)
\end{equation}

\subsection{Two-Particle Extension}
For two distinguishable particles (players A and B), the Hilbert space becomes $\mathcal{H} = \mathcal{H}^A \otimes \mathcal{H}^B$, with the joint state:
\begin{equation}
|\Psi(t)\rangle = \sum_{x_A,x_B} \sum_{s_A,s_B} \Psi_{s_A,s_B}(x_A,x_B,t) |x_A\rangle|s_A\rangle \otimes |x_B\rangle|s_B\rangle
\end{equation}
where $s_A, s_B \in \{R, L\}$.

The non-interacting evolution operator factorizes:
\begin{equation}
U_0 = U^A \otimes U^B = \left[S^A(C^A \otimes I)\right] \otimes \left[S^B(C^B \otimes I)\right]
\end{equation}
with the shift operators:
\begin{equation}
S^i = \sum_{x} \left[ |x+1\rangle\langle x| \otimes |R\rangle\langle R| + |x-1\rangle\langle x| \otimes |L\rangle\langle L| \right], \quad i \in \{A,B\}
\end{equation}

\subsection{Interaction Operator}

Strategic coupling is introduced through a diagonal unitary phase operator acting on the joint position--coin basis,
\begin{equation}
P_{\mathcal{I}} = \exp\!\left[i\,\mathcal{I}(x_A,x_B,s_A,s_B;\theta_A,\theta_B)\right],
\end{equation}
which preserves unitarity while modifying multi-path interference.

For the quantum race studied in this work, we employ a collision-induced interaction phase,
\begin{equation}
\mathcal{I}(x_A,x_B,s_A,s_B;\theta_A,\theta_B)=
\begin{cases}
\pi\cos(\theta_A-\theta_B), & x_A=x_B,\\[4pt]
0, & \text{otherwise}.
\end{cases}
\end{equation}

This choice implements a strategy-dependent scattering phase acquired exclusively upon spatial overlap. The cosine dependence ensures smooth tunability and rotational symmetry in strategy space. Constructive or destructive interference at collision events therefore becomes directly controllable by the players’ strategic parameters.

\subsection{Full Evolution with Interaction}
The complete evolution operator for the interacting system is:
\begin{equation}
U = P_{\mathcal{I}} \cdot U_0 = P_{\mathcal{I}} \cdot \left[ \left(S^A(C^A \otimes I)\right) \otimes \left(S^B(C^B \otimes I)\right) \right]
\end{equation}
The state after $T$ steps is:
\begin{equation}
|\Psi(T)\rangle = U^T |\Psi(0)\rangle
\end{equation}

\subsection{Measurement and Payoff Calculation}
The final measurement is performed in the position basis, yielding the probability distribution:
\begin{equation}
P(x_A, x_B) = \sum_{s_A,s_B} |\langle x_A, s_A, x_B, s_B | \Psi(T) \rangle|^2
\end{equation}
The expected payoff for player $i$ is then:
\begin{equation}
\langle U_i \rangle = \sum_{x_A,x_B} P(x_A, x_B) U_i^{\text{classical}}(x_A, x_B)
\end{equation}
where $U_i^{\text{classical}}$ is the classical payoff function mapping position outcomes to rewards.

\section{Analysis of Nash Equilibria}

\subsection{Non-interacting case}

When $P_{\mathcal{I}}=I$, the evolution factorizes as
\begin{equation}
U_0 = U_0^A\otimes U_0^B,
\end{equation}
and the joint state after $T$ steps remains separable,
\begin{equation}
|\Psi(T)\rangle = |\phi(\theta_A)\rangle\otimes|\chi(\theta_B)\rangle .
\end{equation}

Consequently, the joint distribution satisfies
\begin{equation}
P(x_A,x_B)=p_A(x_A;\theta_A)\,p_B(x_B;\theta_B).
\end{equation}

For the competitive payoff $U_A=x_A-x_B$, the expected utility becomes
\begin{equation}
\langle U_A\rangle = F(\theta_A)-F(\theta_B),
\qquad
F(\theta)=\mathbb{E}[x(T)|\theta].
\end{equation}

The best-response mappings are therefore
\begin{align}
BR_A(\theta_B) &= \arg\max_{\theta_A} F(\theta_A),\\
BR_B(\theta_A) &= \arg\min_{\theta_B} F(\theta_B).
\end{align}

Let $\theta_{\max}$ and $\theta_{\min}$ denote the maximizer and minimizer of $F(\theta)$, respectively. Any Nash equilibrium must satisfy
\begin{equation}
\theta_A^*=\theta_{\max},\qquad \theta_B^*=\theta_{\max},
\end{equation}
since player B can always improve by choosing $\theta_B=\theta_A$ to enforce zero payoff.

Thus the unique equilibrium is the degenerate symmetric profile
\begin{equation}
(\theta_A^*,\theta_B^*)=(\theta_{\max},\theta_{\max}),
\qquad
\langle U_A\rangle=\langle U_B\rangle=0.
\end{equation}

This equilibrium is trivial: strategies are independent and no strategic coupling exists. Genuine competitive structure only emerges once $P_{\mathcal{I}}\neq I$, which breaks separability and introduces interference-mediated strategic dependence.


\begin{thebibliography}{30}

\bibitem{Eisert1999} 
Eisert, J., Wilkens, M., \& Lewenstein, M. (1999). 
Quantum games and quantum strategies. 
\textit{Phys. Rev. Lett.}, 83(15), 3077.

\bibitem{Benjamin2001}
Benjamin, S. C., \& Hayden, P. M. (2001).
Multiplayer quantum games. 
\textit{Phys. Rev. A}, 64(3), 030301.

\bibitem{Marinatto2000}
Marinatto, L., \& Weber, T. (2000).
A quantum approach to static games of complete information. 
\textit{Phys. Lett. A}, 272, 291--303.

\bibitem{Flitney2002}
Flitney, A. P., \& Abbott, D. (2002).
An introduction to quantum game theory.
\textit{Fluctuation and Noise Letters}, 2(04).

\bibitem{Cheon2006}
Cheon, T., \& Tsutsui, I. (2006).
Classical and quantum contents of solvable game theory on Hilbert space.
\textit{Phys. Lett. A}, 348, 147--152.

\bibitem{Khan2013}
Khan, F. S., \& Phoenix, S. J. (2013).
Mini-maximizing two-qubit quantum computations. 
\textit{Quantum Inf. Process.}, 12, 3807--3819.

\bibitem{Frackiewicz2021}
Frąckiewicz, P. (2021).
Non-classical rules in quantum games.
\textit{Entropy}, 23(5), 604.

\bibitem{2016EL11450012}
Deng, X., Deng, Y., Liu, Q., Shi, L., \& Wang, Z. (2016).
Quantum games of opinion formation.
\textit{Europhys. Lett.}, 114, 50012.

\bibitem{Piotrowski2003}
Piotrowski, E. W., \& Sladkowski, J. (2002).
Quantum market games.
\textit{Physica A}, 312, 208--216.

\bibitem{Teeni2023}
Te’eni A, Peled BY, Cohen E, Carmi A  (2023).
Multi-agent dynamical quantum game.
\textit{PLOS ONE}, 18(1), e0280798.

\bibitem{Shi2025}
Shi, H., Zhang, M., Chen, H. et al.  (2025).
Quantum approaches for inference and decision-making in quantum multi-agent frameworks
\textit{Eur. Phys. J. Spec. Top}, 234, 6289–6307.

\bibitem{Du2002}
Du, Jiangfeng and Li, Hui and Xu. et al. (2002).
Experimental Realization of Quantum Games on a Quantum Computer
\textit{Phys. Rev. Lett.}, 88(13), 137902.

\bibitem{Prevedel2007}
Prevedel R, Stefanov A, Walther P, and Zeilinger A (2007).
Experimental realization of a quantum game on a one-way quantum computer
\textit{New J. Phys.}, 9, 205.

\bibitem{Meyer1999}
Meyer, D. A. (1999).
Quantum strategies.
\textit{Phys. Rev. Lett.}, 82, 1052.

\bibitem{Aharonov1993}
Aharonov, Y. and Davidovich, L. and Zagury, N. (1993).
Quantum random walks.
\textit{Phys. Rev. A}, 48(2), 1687-1690.

\bibitem{Ambainis2001}
Ambainis, A.  Bach, E. Nayak, A. Vishwanath, A. Watrous, J. (2001).
One-dimensional quantum walks.
\textit{STOC '01: Proceedings of the thirty-third annual ACM symposium on Theory of computing
}, 37--49.

\bibitem{Venegas2012}
Venegas-Andraca, S. E. (2012).
Quantum walks review.
\textit{Quantum Inf. Process.}, 11, 1015--1106.

\bibitem{Chandrashekar2007}
Chandrashekar, C. M., Srikanth R., and Laflamme R. (2007).
Optimizing quantum walk.
\textit{Phys. Rev. A}, 77, 032326.

\bibitem{Ahlbrecht2012}
Ahlbrecht, A., et al. (2012).
Molecular binding in quantum walks.
\textit{New J. Phys.}, 14, 073050.

\bibitem{Schreiber2012}
Schreiber, A., et al. (2010).
Photons walking the line.
\textit{Phys. Rev. Lett.}, 104, 050502.

\bibitem{Pawel2013}
Kurzyński, P., \& Wójcik, A. (2013).
Quantum walk as measuring device.
\textit{Phys. Rev. Lett.}, 110, 200404.

\bibitem{Preiss2015}
Preiss, P. M., et al. (2015).
Strongly correlated quantum walks.
\textit{Science}, 347, 1229--1233.


\bibitem{Shiratori2025}
Shiratori, H., et al (2025).
Multi-player conflict avoidance through entangled quantum walks
\textit{Phys. Rev. A} 112(6), 062439.

\bibitem{Chen2018}
Solmeyer, N., Dixon, R., \& Balu, R. (2018).
Quantum routing games.
\textit{J. Phys. A}, 51, 455304.



\bibitem{Konno2002}
Konno, N. (2002).
Quantum random walks in one dimension.
\textit{Quantum Inf. Process.}, 1, 345--354.









.

\end{thebibliography}
\end{document}